\documentclass[fleqn,usenatbib]{mnras}

\usepackage{newtxtext,newtxmath}

\usepackage[T1]{fontenc}

\usepackage{ulem}

\DeclareRobustCommand{\VAN}[3]{#2}
\let\VANthebibliography\thebibliography
\def\thebibliography{\DeclareRobustCommand{\VAN}[3]{##3}\VANthebibliography}

\usepackage{graphicx}	
\usepackage{amsmath}	

\def\logR{\ensuremath{\log R^{\prime}_{\mathrm{HK}}}}
\def\loggf{\ensuremath{\log gf}}

\title[ISM effect on transit]{Impact of Mg{\sc ii} interstellar medium absorption on near-ultraviolet exoplanet transit measurements}

\author[A. G. Sreejith et al.]{
A. G. Sreejith$^{1,2}$\thanks{E-mail: sreejith.aickara@oeaw.ac.at},
L. Fossati$^{1}$,
P. E. Cubillos$^{1,3}$,
S. Ambily$^{2}$,
and K. France$^{2}$
\\
$^{1}$Space Research Institute, Austrian Academy of Sciences, Schmiedlstrasse 6, 8042 Graz, Austria\\
$^{2}$Laboratory for Atmospheric and Space Physics, University of Colorado, UCB 600, Boulder, CO, 80309, USA\\
$^{3}$INAF – Osservatorio Astrofisico di Torino, Via Osservatorio 20, 10025 Pino Torinese, Italy
}

\date{Accepted XXX. Received YYY; in original form ZZZ}

\pubyear{2021}

\begin{document}
\label{firstpage}
\pagerange{\pageref{firstpage}--\pageref{lastpage}}
\maketitle

\begin{abstract}
Ultraviolet (UV) transmission spectroscopy probes atmospheric escape, which has a significant impact on planetary atmospheric evolution. If unaccounted for, interstellar medium absorption (ISM) at the position of specific UV lines might bias transit depth measurements, and thus potentially affect the (non-)detection of features in transmission spectra. Ultimately, this is connected to the so called ``resolution-linked bias'' (RLB) effect. We present a parametric study quantifying the impact of unresolved or unconsidered ISM absorption in transit depth measurements at the position of the Mg{\sc ii}\,h\&k resonance lines (i.e. 2802.705\,\AA\ and 2795.528\,\AA, respectively) in the near-ultraviolet spectral range. We consider main-sequence stars of different spectral types and vary the shape and amount of chromospheric emission, ISM absorption, and planetary absorption, as well as their relative velocities. We also evaluate the role played by integration bin and spectral resolution. We present an open-source tool enabling one to quantify the impact of unresolved or unconsidered Mg{\sc ii} ISM absorption in transit depth measurements. We further apply this tool to a few already or soon to be observed systems. On average, we find that ignoring ISM absorption leads to biases in the Mg{\sc ii} transit depth measurements comparable to the uncertainties obtained from the observations published to date. However, considering the bias induced by ISM absorption might become necessary when analysing observations obtained with the next generation space telescopes with UV coverage (e.g. LUVOIR, HABEX), which will provide transmission spectra with significantly smaller uncertainties compared to what obtained with current facilities (e.g. HST).
\end{abstract}



%
\section{Introduction}
Atmospheric escape is one of the key processes affecting the long-term evolution of a planetary atmosphere and it plays a pivotal role in shaping the currently observed exoplanet population \citep[e.g.][]{Lecavelier2007,lopez2013,owen2017,jin2018,kubyshkina2018}. The first detection of exoplanet atmospheric escape occurred through Ly$\alpha$ transmission spectroscopy \citep{vidal2003}. Since then, this technique has been central in observationally probing upper atmospheres and escape \citep[e.g.][]{lecavelier2012,ehrenreich2015,bourrier2018}. One of the most significant problems of Ly$\alpha$ transmission spectroscopy is that, because of interstellar medium (ISM) absorption, Ly$\alpha$ transit observations are possible only for nearby stars, typically within 50\,pc \citep[e.g.][]{ehrenreich2012}. Furthermore, even for stars for which Ly$\alpha$ transmission spectroscopy is possible, the observations typically probe high velocity neutral hydrogen, because the core of the line, which is where most of the planetary absorption is located, is often (fully) absorbed by neutral hydrogen in the ISM  \citep{Guinan2016,Bourrier2017}.

Transmission spectroscopy observations aiming at studying atmospheric escape have also been conducted at near-ultraviolet (NUV) wavelengths. This is to gain from the higher NUV stellar flux, compared to the far-ultraviolet (FUV), but also to gain from the fact that in the NUV the stellar emission is more temporally stable and spatially homogeneous \citep{haswell2010,haswell2012}. Just a few exoplanets have been observed in the NUV from space using either spectroscopy or broad-band photometry \citep{fossati2010,haswell2012,vidal2013,salz2019,sing2019,cubillos2020,wakeford2020,Lothringer2022}. The NUV spectral range is dominated by a plethora of strong lines of several metals, with the most dominant being Mg and Fe. The most prominent NUV spectral features are the Mg{\sc ii}\,h\&k resonance lines at $\approx$2800\,\AA. In late-type stars, these lines present a core reversal (i.e. emission) that originates in the chromosphere and transition region  \citep[e.g.][]{hall2008,youngblood2022}. Because these are resonance lines, Mg is a rather abundant element, and singly ionised Mg is the most abundant Mg ion in the ISM \citep[e.g.][]{frisch}, stellar emission at the core of these lines is affected by ISM absorption, similarly to Ly$\alpha$, though typically less pronounced. Indeed, ISM Mg{\sc ii} absorption lines have been ubiquitously found in the NUV spectra of stars, regardless of their distance from Earth \citep[e.g.][]{redfield2002}.

Similar to the Ly$\alpha$ line, in systems hosting transiting planets with an extended atmosphere, depending on the radial velocity of the ISM cloud with respect to that of the star, the Mg{\sc ii} ISM absorption features may lie on top of the planetary absorption features. If this ISM absorption, which occurs after the planetary atmosphere has imprinted its signature in the stellar flux, is unaccounted for, the transmission spectrum at the core of the Mg{\sc ii} lines might be biased. This is the case of broad band photometry, but it may occur also with transmission spectroscopy, for example if the low signal-to-noise ratio of the data requires binning and/or does not enable one to resolve the ISM absorption features. As shown below (Section~\ref{sec:analysis}), this occurs as a result of the different shape of the stellar, planetary absorption, and ISM absorption features, despite that of the transit depth is a differential measurement and the impact of ISM absorption cancels out when considering a single wavelength point. The origin of this phenomenon is similar to the one that \citet{Drake2017} called ``resolution-linked bias'' (RLB).

The most representative case is certainly that of the WASP-12 system. The late F-type star WASP-12 hosts a hot Jupiter with an orbital period of about 1\,day \citep{hebb2009}. \citet{fossati2010} , \citet{haswell2012} and \citet{Nichols2015} presented the results of HST/COS NUV transmission spectroscopy observations of WASP-12b that also covered the Mg{\sc ii}\,h\&k lines. The observations indicated that the planet hosts an escaping atmosphere that is rich in metals as heavy as iron. The COS spectra revealed that, surprisingly, the core of the Mg{\sc ii} lines does not show any reversal, which has been explained by a combination of ISM absorption and absorption local to the system caused by a cloud of planetary escaped material surrounding the star \citep{haswell2012,fossati2013,debrecht2018,dwivedi2019}. Furthermore, the large distance to the star \citep[$\approx$430\,pc;][]{gaiadr2} implied that, to gather enough signal for significantly detecting the planet and its atmospheric signature, the data needed to be heavily rebinned, integrating across about 20\,\AA\ on either side of the Mg{\sc ii} lines. Therefore, the transit depth measured in the region of the Mg{\sc ii}\,h\&k lines might have been biased by the unresolved ISM absorption.

WASP-121b and HD209458b have also been observed employing NUV HST/STIS transmission spectroscopy covering the Mg{\sc ii}\,h\&k lines  \citep{sing2019,vidal2013,cubillos2020}. Despite the higher signal-to-noise of the data, compared to that of WASP-12b, which enabled resolving the ISM absorption features on top of the chromospheric stellar emission, the analysis had been carried out integrating the flux in the line cores without excluding the regions affected by the ISM absorption. This may have biased the transmission spectra around the line cores.

There is therefore the need to identify, and eventually quantify, the impact of unresolved or unconsidered ISM absorption in the core of the Mg{\sc ii} lines in transit depth measurements. 
To this end, we employ synthetic spectra and take into account a number of stellar, planetary, ISM, instrumental, and analysis parameters. Furthermore, we present an open-source code enabling one to estimate the correction to the transit depth caused by the unresolved or unconsidered ISM absorption. This work is organised as follows. In Sections~\ref{sec:params} and \ref{sec:analysis}, we present the input parameters and the analysis methods. In Section~\ref{sec:results}, we show the results, which we discuss in Section~\ref{sec:discussion}. We gather the conclusions in Section~\ref{sec:conclusion}.
\section{Input parameters}\label{sec:params}
%
\begin{figure*}
\begin{center}
\resizebox{\hsize}{!}
{\includegraphics[width=0.5\textwidth]{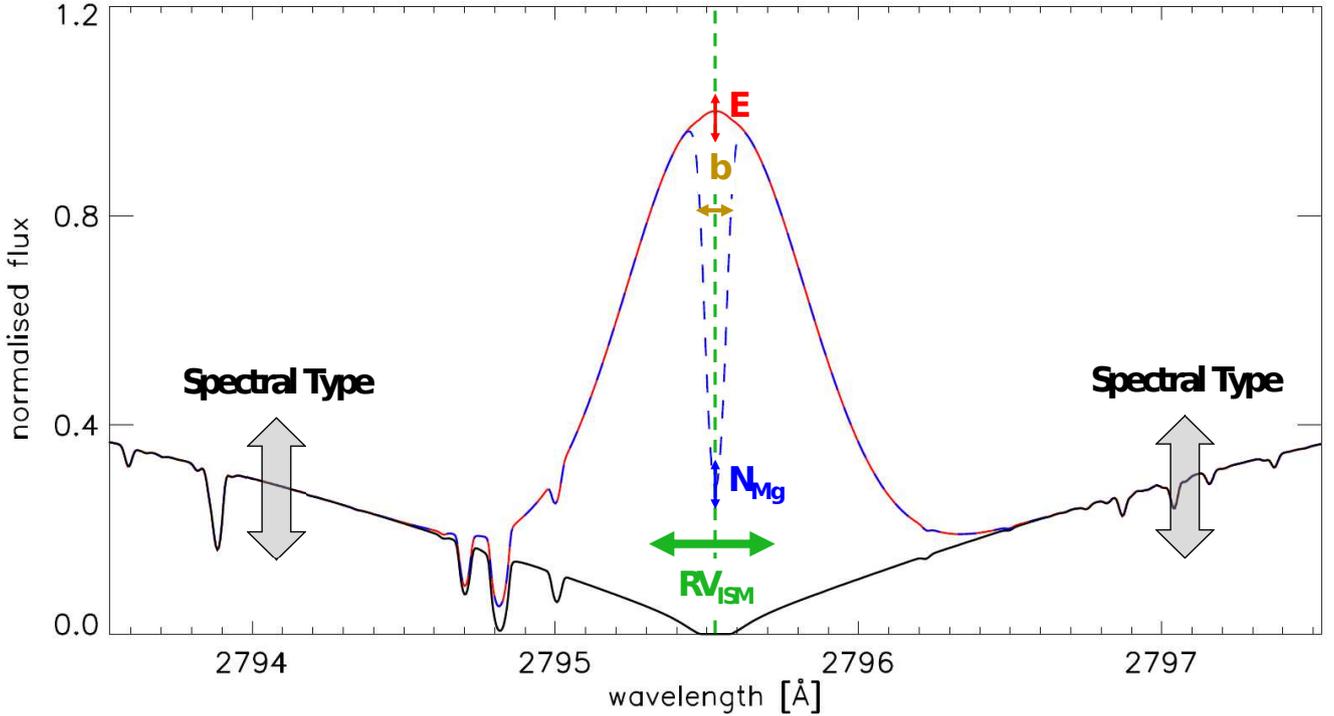}}
\caption{Synthetic photospheric spectrum of a Sun-like star around the core of the Mg{\sc ii}\,k line (black solid line). The red line shows the synthetic Mg{\sc ii} chromospheric line core emission, constructed as described in the text and considering a total Mg{\sc ii} chromospheric emission of 200\,erg\,cm$^{-2}$\,s$^{-1}$. The dashed blue line shows the ISM absorption feature, placed at the line core and computed considering an ISM Mg{\sc ii} column density of 10$^{12}$\,cm$^{-2}$ and an ISM broadening $b$-parameter of 4\,km\,s$^{-1}$. The main parameters affecting the line shapes are labelled, while the arrows indicate how these parameters modify the line profiles.}
\label{fig:fig1}
\end{center}
\end{figure*}

We employ synthetic spectra to quantify the impact of unresolved ISM absorption at the position of the Mg{\sc ii}\,h\&k resonance lines in planetary transit measurements. We build up the synthetic spectra by adding the ISM and planetary absorption and line emission features on top of the photospheric spectra, thus enabling us to control each relevant physical parameter. Furthermore, we consider also instrumental and analysis parameters, namely spectral resolution and integration bin. The different ingredients are added in sequence following what occurring in reality, that is we start with the photospheric spectrum to which we add in order the contribution of the chromospheric emission, the planetary absorption, and the ISM absorption. Finally, we convolve the resulting spectrum by the instrumental spectral resolution and eventually bin the data. Each ingredient and the applied operations are described here below and in Section~\ref{sec:analysis}.
\subsection{Stellar and ISM parameters}
The stellar parameters control the photospheric emission and the Mg{\sc ii}\,h\&k line core emission. We computed stellar photospheric spectra employing the LLmodels stellar atmosphere code \citep{Shulyak2004}, which assumes local thermodynamical equilibrium (LTE). In the region covered by the Mg{\sc ii}\,h\&k lines, the input photospheric spectra have a spectral resolution of 560,000 and a wavelength sampling of 0.005\,\AA. We computed the photospheric spectra considering a range of effective temperatures (3500\,$<$\,$T_{\rm eff}$\,$<$\,10000\,K, in steps of 100\,K), a $\log{g}$ value of 4.4 (i.e. main-sequence stars), and solar composition \citep{asplund2009}. Having fixed surface gravity and composition, the wings of the photospheric Mg{\sc ii}\,h\&k lines are controlled exclusively by $T_{\rm eff}$ (Figure~\ref{fig:fig1}). 

We added the stellar chromospheric emission on top of the photospheric Mg{\sc ii}\,h\&k lines following the algorithm described by \citet{fossati2017a} and \citet{sreejith2019}. In short, we obtained the stellar radius ($R_{\rm s}$) on the basis of $T_{\rm eff}$ by interpolating Table~5\footnote{{\tt http://www.pas.rochester.edu/$\sim$emamajek/\\EEM\_dwarf\_UBVIJHK\_colors\_Teff.txt}} of \citet{pecaut2013}. The strength of the line core emission is controlled by the $E$ parameter (Figure~\ref{fig:fig1}), which corresponds to the disk integrated Mg\,{\sc ii}\,h\&k chromospheric emission flux at 1\,AU in units of erg\,cm$^{-2}$\,s$^{-1}$. This parameter can be estimated for example from the \logR~ value (see Section~\ref{sec:mg2emission}). Assuming the gas responsible for the chromospheric emission is optically thin, we divide this emission between the two Mg\,{\sc ii} lines according to their oscillator strengths taken from the VALD database \citep{piskunov,kupka,ryab}, which leads to a line strength ratio of about two. We remark that non-local thermodynamic equilibrium (NLTE) effects, which we do not consider, may slightly alter the division of the chromospheric emission flux between the two lines based on the sole oscillator strengths, but we find that this does not have a noticeable impact on the results. For the width of the emission lines, we employed values derived from a high-resolution HST/STIS spectrum of $\alpha$\,Cen\,A \citep[as in][]{sreejith2019}, namely a full width at half maximum (FWHM) of 0.518\,\AA\ and 0.546\,\AA\ for the Mg{\sc ii}\,k and Mg{\sc ii}\,h lines, respectively.  However, the width and shape of the emission lines varies slightly as a function of spectral type and luminosity \citep[e.g.][]{wilson1957,ayres1979,redfield2002,gunar2021} and we test the impact of varying the width of the emission lines in Section~\ref{sec:emission_width}.

We modelled the ISM absorption using a Voigt profile. The ISM line absorption is controlled by the Mg\,{\sc ii} column density ($N_{\rm MgII}$), the broadening parameter ($b$), and the ISM radial velocity with respect to that of the star ($RV_{\rm ISM}$). These three parameters control the strength, width, and position of the ISM absorption features, respectively. The range of values describing the ISM absorption we consider for these parameters have been taken on the basis of the results of \cite{redfield2002}, who extracted Mg\,{\sc ii} ISM parameters from NUV spectra of stars lying within 100\,pc. To maximise the impact of the ISM absorption on the emission lines we adopt $RV_{\rm ISM}$\,=\,0\,km\,s$^{-1}$, but also test how the results change by varying the parameters controlling the ISM absorption (see Section~\ref{sec:ism_parameters}).

Figure~\ref{fig:fig1} shows the impact that each of the input parameters has on the shape and strength of the stellar and ISM line profiles. The stellar effective temperature modifies the strength of the photospheric Mg{\sc ii} line wings. The Mg\,{\sc ii} emission flux ($E$) increases or decreases the chromospheric flux. The Mg\,{\sc ii} column density determines the strength of the ISM absorption feature, while the $b$-parameter controls the width of the ISM line profile. The ISM radial velocity controls the position of the ISM absorption line with respect to that of the Mg{\sc ii}\,h\&k line cores in the reference frame of the star.
\subsection{Planetary parameters}\label{sec:2-plparam}
We consider a simplified planetary transmission spectrum in which the transit depth (i.e. the square of the planet-to-star radius ratio; R$_{\rm p}^2$/R$_{\rm s}^2$) is fixed at 1\%  (i.e. a typical hot Jupiter orbiting a Sun-like star), except for the region covered by the Mg{\sc ii}\,h\&k lines, where we consider that the peak of the planetary absorption at the position of the Mg{\sc ii}\,k line leads to a transit depth of either 5\% or 10\% \citep{sing2019,fossati2021}. We further consider that the Mg{\sc ii} planetary absorption divides between the two lines following the ratio of the line oscillator strengths. This implies the assumption that the planetary atmospheric gas producing the absorption is optically thin, which is not necessarily the case. Indeed, for example the theoretical Mg{\sc ii}\,h\&k absorption obtained for the ultra-hot Jupiter KELT-9b considering NLTE effects is about the same for the two lines \citep{fossati2021}. However, the relative strength of the two lines depends on the physical characteristics of the considered planetary atmosphere, particularly as a result of NLTE effects. We are therefore bound to take an assumption, that we choose being that of optically thin gas, but show in Section~\ref{sec:impact_pl_star} that this assumption has a negligible impact on the results.

\begin{figure*}
\begin{center}
\resizebox{\hsize}{!}
{\includegraphics[width=\textwidth]{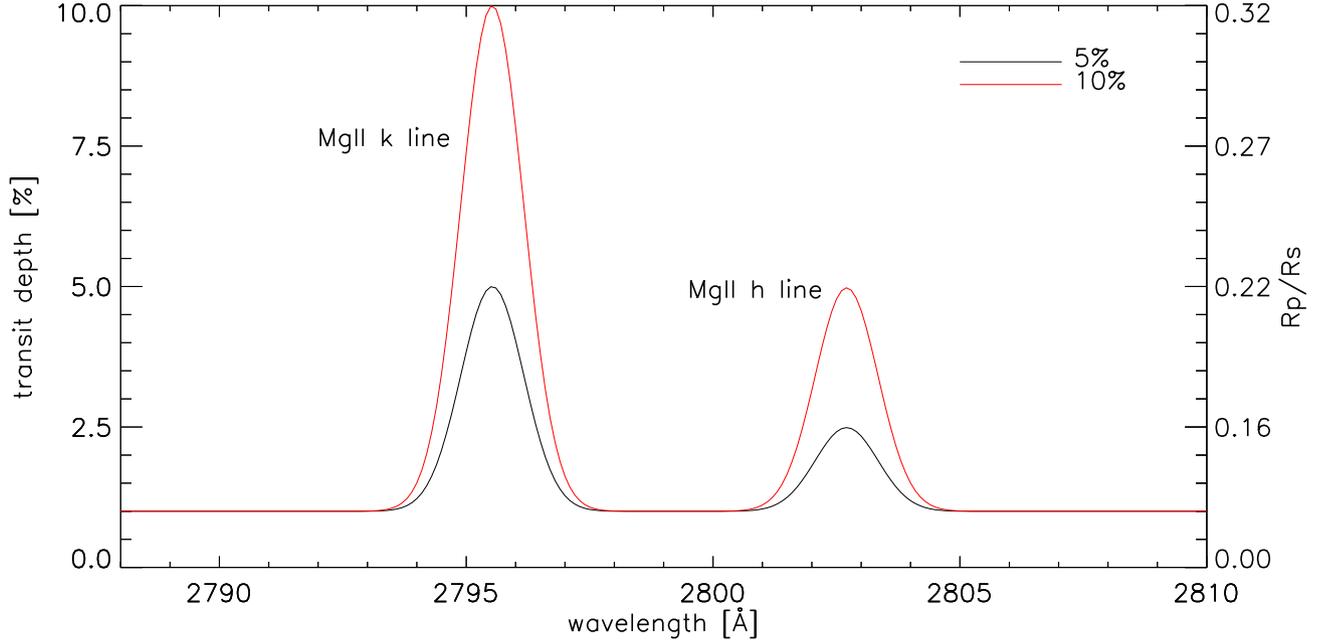}}
\caption{Synthetic planetary transmission spectra computed assuming a constant transit depth of 1\% and peak absorption at the position of the Mg{\sc ii}\,k line of 5\% or 10\%. }
\label{fig:figtds}
\end{center}
\end{figure*}

We modelled the planetary absorption features at the position of the Mg{\sc ii}\,h\&k lines as Gaussians having a FWHM of 1.5\,\AA. This value has been guided by the results of \cite{sing2019}, but it can vary significantly from planet to planet, which is why in the following we look also at the impact of the planetary absorption line width on the results. We remark that we do not account for limb darkening, therefore (R$_{\rm p}$/R$_{\rm s}$)$^2$ corresponds to the transit depth. This is because we are interested in the transit depth difference between measurements obtained with and without ISM absorption for which the effect of limb darkening cancels out. Figure~\ref{fig:figtds} shows the planetary transmission spectrum for both considered cases in which the peak of the absorption at the position of the Mg{\sc ii}\,k line amounts to 5\% or 10\%.
%
\subsection{Instrument and analysis parameters} 
The instrumental parameter we consider in this work is the spectral resolution ($R$). In particular, we employ three spectral resolution values of 114\,000, 30\,000 and 2\,500, which correspond to the cases of ultra-high resolution spectrographs on-board HST (i.e. $R$\,=\,114\,000; STIS E230H) and high resolution spectrographs on-board HST \citep[$R$\,=\,30\,000; STIS E230M;][]{green2012}, and low resolution spectrographs on board SmallSats such as CUTE \citep[$R$\,=\,2\,500;][]{fleming2018}. The analysis parameter we consider is the integration bin used for calculating transit depths.
\section{Analysis}\label{sec:analysis}
%
\begin{figure*}
\begin{center}
\resizebox{\hsize}{!}
{\includegraphics[width=\textwidth]{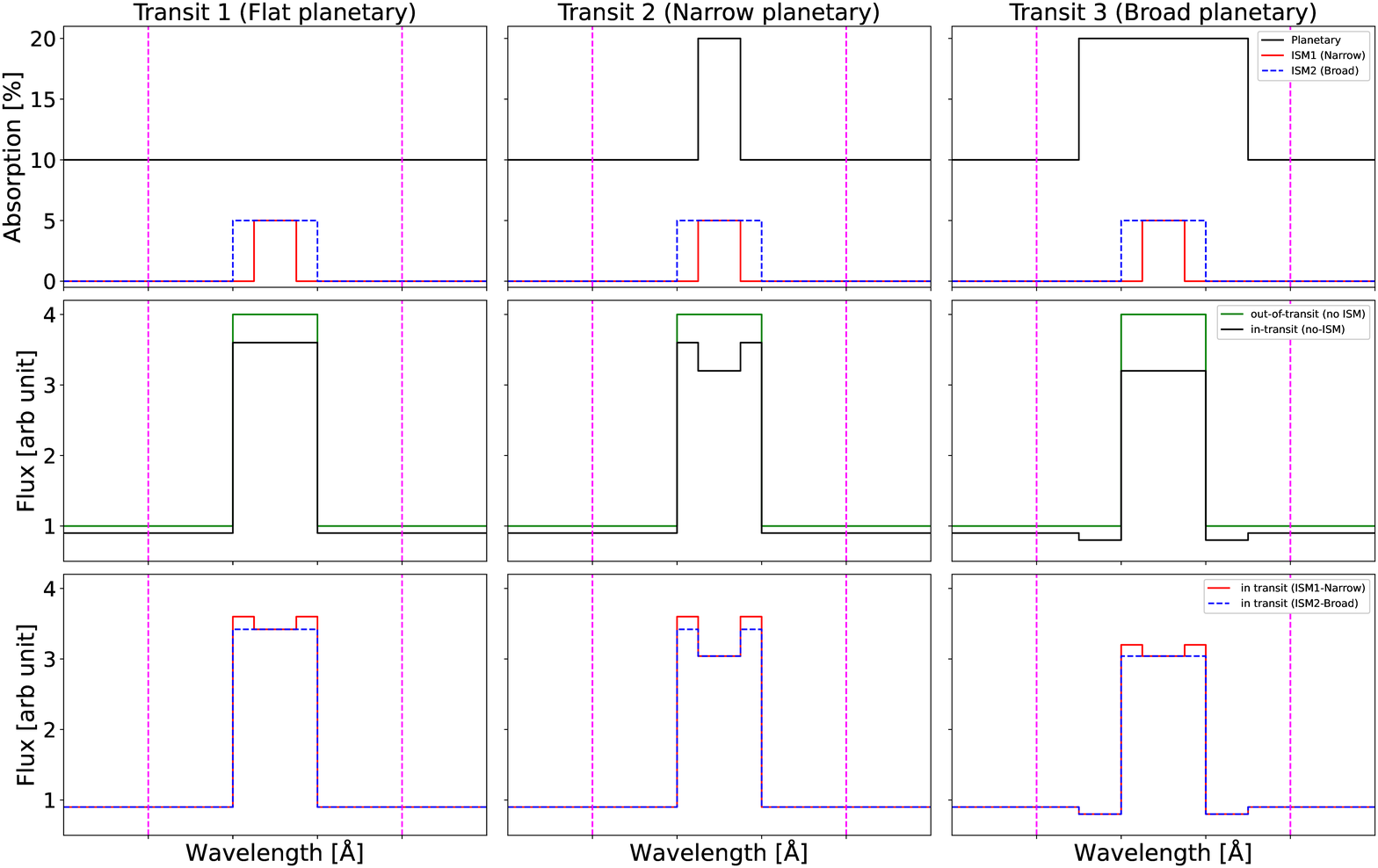}}
\caption{Schematic of the effect of rectangular ISM and planetary absorption features of different strengths and widths on a rectangular stellar emission located at the position of the Mg{\sc ii}\,k line. Top: Considered planetary absorption (black; left y-axis) and ISM absorption features (red and blue; right y-axis) as a function of wavelength. From left to right, the planetary absorption features are constant (i.e. not wavelength dependent), narrow, and broad, while within each panel the ISM absorption features are either narrow (red) or broad (blue). Middle: Stellar emission line with (black) and without (green) planetary absorption. In both cases, there is no ISM absorption. Bottom: Stellar emission line with planetary and narrow (red) and broad (blue) ISM absorption. The vertical magenta lines indicate the integration bin considered for the computation of the line fluxes listed in Table~\ref{table:1}.}
\label{fig:egfig}
\end{center}
\end{figure*}

In general, the radiative transfer equation for pure absorption can be written as
\begin{equation}
dI = -I d\tau\,,
\end{equation}
where $I$ is the intensity of the light passing through an absorbing gas of optical depth $\tau$ and $dI$ represents the absorption. In the case of stellar radiation passing through the ISM, the intensity following ISM absorption is
\begin{equation}
I = I_0 e^{-\tau_{\rm ISM}}\,,
\end{equation}
where $I_0$ is the stellar emission, $I$ the stellar radiation following ISM absorption, and $\tau_{\rm ISM}$ the optical depth of the ISM cloud absorbing the stellar light. 
The stellar flux observed at Earth in- and out of transit is

\begin{equation}
F({\rm ISM})_{\rm in/out} = F_{\rm in/out} e^{-\tau_{\rm ISM}}\,.
\end{equation}

Therefore, the variation of the observed stellar flux as a result of ISM absorption in- and out of transit is

\begin{equation}
\Delta_{\rm in/out} = F_{\rm in/out} - F({\rm ISM})_{\rm in/out} = F_{\rm in/out}\,(1-e^{-\tau_{\rm ISM}})\,.
\end{equation}

Here, $\Delta_{\rm out}$\,$>$\,$\Delta_{\rm in}$, because $F_{\rm out}$\,$>$\,$F_{\rm in}$. However, the transit depth ($d_0$) is a differential measurement, which in case of no ISM absorption is
\begin{equation}
d_0 = \frac{F_{\rm out} - F_{\rm in}}{F_{\rm out}} = 1 - \frac{F_{\rm in}}{F_{\rm out}}\,.
\end{equation}
In presence of ISM absorption, the transit depth ($d_{\rm ISM}$) is
\begin{equation}
d_{\rm ISM}= 1 - \frac{F({\rm ISM})_{\rm in}}{F({\rm ISM})_{\rm out}} = 1 - \frac{F_{\rm in}\,(1-e^{-\tau_{\rm ISM}})}{F_{\rm out}\,(1-e^{-\tau_{\rm ISM}})} = d_0\,,
\end{equation}
demonstrating that a monochromatic transit depth measurement is not affected by ISM absorption.

However, this applies to a single wavelength range at infinite spectral resolution and in the ideal case that all involved line profiles have the same shape. Here, we employ the simplified case of rectangular spectral lines (to enable intuitive line integration) to demonstrate that in reality ISM absorption does affect transit depth measurements as a result of different line shapes in combination with a finite spectral resolution. Figure~\ref{fig:egfig} shows a step-by-step construction of three ideal rectangular stellar emission line profiles accounting for different planetary and ISM absorption features, also of rectangular shape. Each of the top panels show three different profiles, one for the planetary absorption (from left to right: constant, narrow, and broad) and two for the ISM absorption (narrow and broad). The middle panels show the stellar emission line profiles before and after planetary absorption, without ISM absorption. The bottom panels show the stellar line profiles following absorption from both planet and ISM. Each panel shows also the integration bin taken into consideration. The top section of Table~\ref{table:1} shows the integrals of the line fluxes in all possible considered cases of ISM and planetary absorption, while the bottom section of Table~\ref{table:1} shows the transit depths obtained from combining the line fluxes. In case of a constant (i.e. not wavelength dependent) planetary absorption, ISM absorption has no influence on the transit depth, but this is not the case when the planetary absorption is wavelength dependent. Furthermore, in case of the narrow planetary absorption feature the deviation from the ISM-free transit depth is greater for the narrow ISM absorption feature, while in case of the broad planetary absorption feature the deviation from the ISM-free transit depth is greater for the broad ISM absorption feature. This demonstrates that ISM absorption can influence transit depth measurements and that this influence depends on the specific shapes of the stellar, ISM, planetary features, spectral resolution, and integration bin in a complicated manner.
\begin{table}
\caption{Integrated line fluxes obtained from all combinations of ISM and planetary absorption features considered in Figure~\ref{fig:egfig} (top half of the table) and relative transit depth values (bottom half of the table).}     
\label{table:1}
\centering                                      
\begin{tabular}{c c c c c}          
\hline\hline                        
& No & Flat & Narrow & Broad \\    
& planetary & planetary & planetary & planetary \\    
& absorption & absorption  & absorption & absorption \\    
\hline                                  
    No ISM     & 6.0 & 5.40 & 5.20 & 4.90   \\ 
    ISM narrow & 5.9 & 5.31 & 5.12 & 4.82       \\
    ISM broad  & 5.8 & 5.22 & 5.03 & 4.74    \\
 \hline                                            
    $d_0$                & - & 0.1000 & 0.1333 & 0.1833   \\ 
    $d_{\rm ISM,narrow}$ & - & 0.1000 & 0.1322 & 0.1831       \\
    $d_{\rm ISM,broad}$  & - & 0.1000 & 0.1328 & 0.1828    \\
\hline                                            
\end{tabular}
\end{table}
We describe in detail the modeling approach, which aims at reproducing as close as possible the physical and instrumental processes occurring in transit depth measurements obtained from transmission spectroscopy observations covering the Mg{\sc ii}\,h\&k lines. With the parameters described in Section~\ref{sec:params}, we compute the stellar flux accounting for chromospheric emission, but without ISM absorption ($F(\lambda)$) 
as
\begin{equation}
F(\lambda) = F(\lambda)_{\rm photo} + F(\lambda)_{\rm chromo}\,,
\end{equation}
where $F(\lambda)_{\rm photo}$ is the stellar photospheric flux and $F(\lambda)_{\rm chromo}$ is the Gaussian-shaped Mg\,{\sc ii} chromospheric emission. We then obtain the stellar flux during transit ($F_{\rm p}(\lambda)$) by multiplying the stellar flux to the wavelength-dependent planetary absorption 
\begin{equation}
F_{\rm p}(\lambda) = F(\lambda)\,\,\left[1-d(\lambda)\right]\,.
\end{equation}

Then, the in-transit and out-of-transit fluxes including ISM absorption are given by

\begin{equation}
F_{\rm p, ISM}(\lambda) = F_{\rm p}(\lambda) {\cal T}_{\rm ISM}(\lambda)
\end{equation}
\begin{equation}
F_{\rm ISM}(\lambda) = F(\lambda)\,\,\rm {\cal T}_{\rm ISM}(\lambda)\,,
\end{equation}
where ${\cal T}_{\rm ISM}(\lambda)$ is the function simulating the effect of ISM Voigt profiles located at the position of the Mg{\sc ii}\,h and Mg{\sc ii}\,k lines.

The fluxes described above are then convolved with the instrument spectral resolution profile to obtain the observed fluxes, that is


\begin{equation}
F_{\rm IR}(\lambda)=F(\lambda)\,\ast{\rm IR}\,, 
\end{equation}
\begin{equation}
F_{\rm p,IR}(\lambda) = F_{\rm p}(\lambda)\,\ast{\rm IR}\,,
\end{equation}
\begin{equation}
F_{\rm ISM,IR}(\lambda)=F_{\rm ISM}(\lambda)\,\ast{\rm IR}\,, 
\end{equation}
and
\begin{equation}
F_{\rm p,ISM,IR}(\lambda) = F_{\rm p,ISM}(\lambda)\,\ast{\rm IR}\,,
\end{equation}
where IR is the Gaussian shaped instrumental resolution and $\ast$ is the convolution symbol. We further integrate the flux in wavelength bins (see Section~\ref{sec:results}) and compute the transit depth per integration bin without and with ISM absorption, respectively, as 
\begin{equation}
d_0 = 1 - \frac{F_{\rm p,IR}(\lambda)}{F_{\rm IR}(\lambda)}
\end{equation}
and 
\begin{equation}
d_{\rm ISM} = 1 - \frac{F_{\rm ISM,p,IR}(\lambda)}{F_{\rm ISM,IR}(\lambda)}\,.
\end{equation}
We finally extract the impact of ISM absorption on the transit depth by computing the difference between $d_{\rm ISM}$ and $d_0$ (i.e. $\Delta d$\,=\,$d_{\rm ISM}$-$d_0$) for a given integration bin as a function of the input parameters. Following Table~\ref{table:1}, $\Delta d$ should be negative. The $\Delta d$ parameter is a measure of the influence of not considering the impact of Mg{\sc ii} ISM absorption in transit depth measurements covering the Mg{\sc ii}\,h\&k lines. At infinite resolution and at situations not considering the instrument broadening and integration bins the value of $\Delta d$ would be zero, but when  instrument resolution and integration bins comes into play $\Delta d$ has a non-zero value indicating that this effect is geometrical in nature caused by differences in shape and strength of the chromospheric emission, the planetary absorption, and the interstellar absorption features, in a similar fashion to that of the RLB effect \citep{Drake2017}. To enable a generalisation of the results, we ignore a range of second-order physical phenomena, such as the motion of planetary absorption features during a transit \citep[e.g.][]{Snellen2010}, the (atmospheric) Rossiter-McLaughlin effect \citep[e.g.][]{Borsa2019}, center-to-limb variations \citep[e.g.][]{Yan2017}, stellar spots and flares affecting line emission, and limb darkening.
\section{Results}\label{sec:results}
%
\begin{figure*}
\begin{center}
\resizebox{\hsize}{!}
{\includegraphics[width=\textwidth]{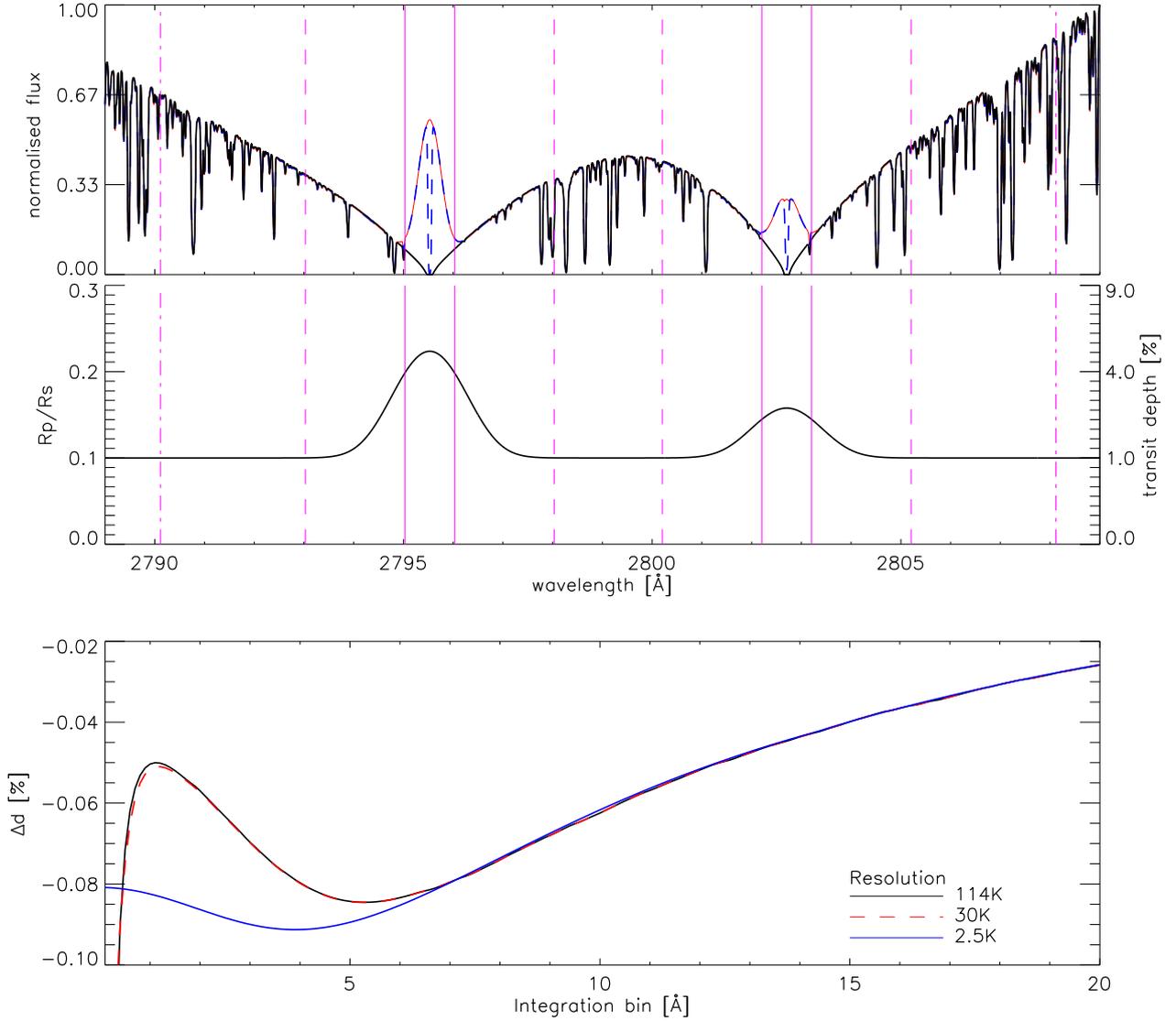}}
\caption{Top: Spectrum of a Sun-like star in the region of the Mg\,{\sc ii}\,h\&k lines (black solid line). The red solid line represents a Mg{\sc ii} line core emission of 100\,erg\,cm$^{-2}$\,s$^{-1}$, while the blue dashed line shows the ISM absorption features obtained considering $\log{N_{\rm MgII}}$\,=\,12.5, $b$\,=\,3\,km\,s$^{-1}$, and $RV_{\rm ISM}$\,=\,0\,km\,s$^{-1}$. Middle: Planetary transmission spectrum for a peak absorption at the position of the Mg{\sc ii}\,k line of 5\%. Bottom: Transit depth difference ($\Delta d$\,=\,$d_{\rm ISM}-d_0$) as a function of integration bin, in \AA, for the three considered spectral resolutions of $R$\,=\,114000 (black), $R$\,=\,30000 (red dashed), and $R$\,=\,2500 (blue). In the top and middle panels, the vertical magenta lines indicate the edges of three different integration bins that are marked in the bottom panel following the same line style (see also Section~\ref{sec:results})}.
\label{fig:fig2}
\end{center}
\end{figure*}
\begin{figure*}
\begin{center}
\resizebox{\hsize}{!}
{\includegraphics[width=\textwidth]{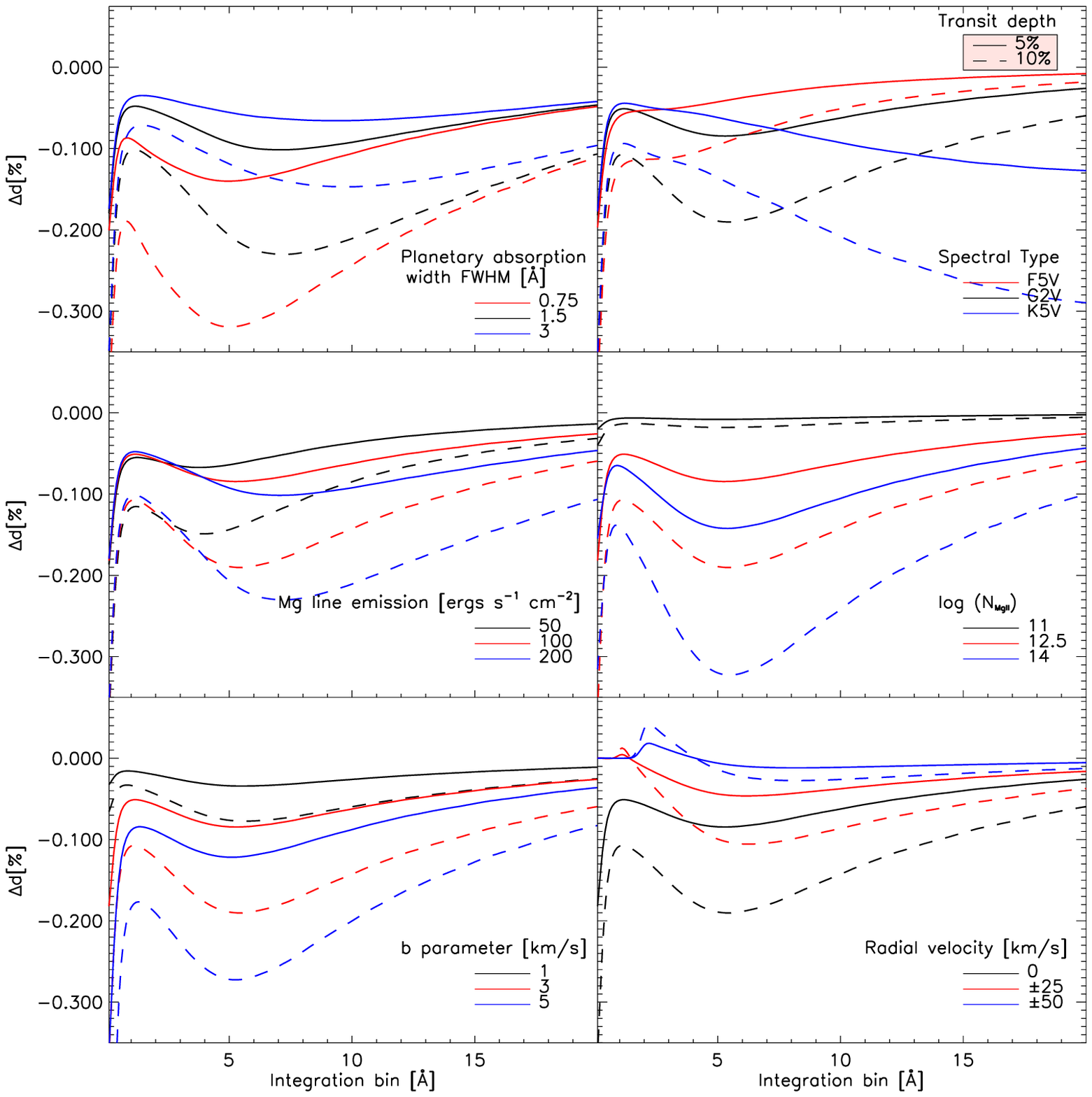}}
\caption{$\Delta d$ as a function of integration bin for two values of the peak planetary absorption at the position of the Mg{\sc ii}\,k line (5\%, solid lines; 10\%, dashed lines) and for different FWHM of the Mg{\sc ii} planetary absorption features (top-left), stellar spectral types (top-right), $E$ values (middle-left), $\log{N_{\rm MgII}}$ values (middle-right), $b$-parameter values (bottom-left), and $RV_{\rm ISM}$ (bottom-right) values, as indicated by the legend located in each panel.}
\label{fig:fig3}
\end{center}
\end{figure*}

We present here the impact of varying the input parameters on $\Delta d$ to identify the effect of unresolved or unconsidered Mg{\sc ii} ISM absorption in transit depth measurements and transmission spectra. Unless otherwise stated, all results presented below have been obtained considering a Sun-like star (i.e. $T_{\rm eff}$\,=\,5800\,K; $R_{\rm s}$\,=\,1\,$R_{\odot}$). Furthermore, in the following when not varying a certain input parameter, we used the default values that are $E$\,=\,100\,erg\,cm$^{-2}$\,s$^{-1}$, $\log{N_{\rm MgII}}$\,=\,12.5, $b$\,=\,3\,km\,s$^{-1}$, $RV_{\rm ISM}$\,=\,0\,km\,s$^{-1}$, $R$\,=\,30000, an integration bin of 5\,\AA, and a width of the planetary absorption features of 1.5\,\AA.

As the transit depth measurement depends on the integration window, our first analysis involved varying the integration bin for different input parameters. Figure~\ref{fig:fig2} illustrates the definition of integration bin considered in this work. The integration bin corresponds to the sum of two equal bins centered on each of the Mg{\sc ii} lines. For example, an integration bin of 2\,\AA\ (solid magenta vertical lines in Figure~\ref{fig:fig2}) corresponds to two 1\,\AA\ wide bins centred around each of the Mg{\sc ii} lines. To avoid discontinuities, when the integration bin becomes large enough that the two bins centred on each line start overlapping (at $\approx$7.2\,\AA\ bin around each line), the integration bin becomes a single bin centred in the middle of the Mg{\sc ii} lines (e.g. dash-dotted magenta vertical lines in Figure~\ref{fig:fig2}).
\subsection{Impact of spectral resolution}
The bottom panel of Figure~\ref{fig:fig2} shows how $\Delta d$ varies as a function of integration bin, for the three considered spectral resolution values. We find that ignoring ISM absorption leads to underestimating the transit depth. At high spectral resolution and for small integration bins, |$\Delta d$| is large, while with increasing integration width |$\Delta d$| first decreases sharply as a result of considering more stellar flux unaffected by the ISM absorption, and then increases gently until it reaches a maximum, which occurs at a position corresponding to the integration bin that covers exactly the whole of the Mg{\sc ii} planetary absorption feature. Beyond this point, $|\Delta d|$ decreases gently towards zero with increasing integration bin. At lower spectral resolution, the behaviour is similar, except for a significant flattening of the $|\Delta d|$ variations as a function of integration bin. Figure~\ref{fig:fig2} indicates that the spectral resolution plays a role at small integration bins, while at large integration bins the spectral resolution has no noticeable effect on $\Delta d$. With the considered set of parameters, the impact of ignoring ISM absorption at the core of the Mg{\sc ii}\,h\&k lines in transit depth measurements is typically small for integration bins that contain the whole of the ISM absorption features, and does not reach the 0.1\% level, that is below the typical uncertainties of NUV transit depth measurements \citep[see e.g.][]{fossati2010,haswell2012,sing2019,cubillos2020}. 
\subsection{Impact of planetary and stellar parameters}\label{sec:impact_pl_star}
\begin{figure}
\begin{center}
\includegraphics[width=\hsize]{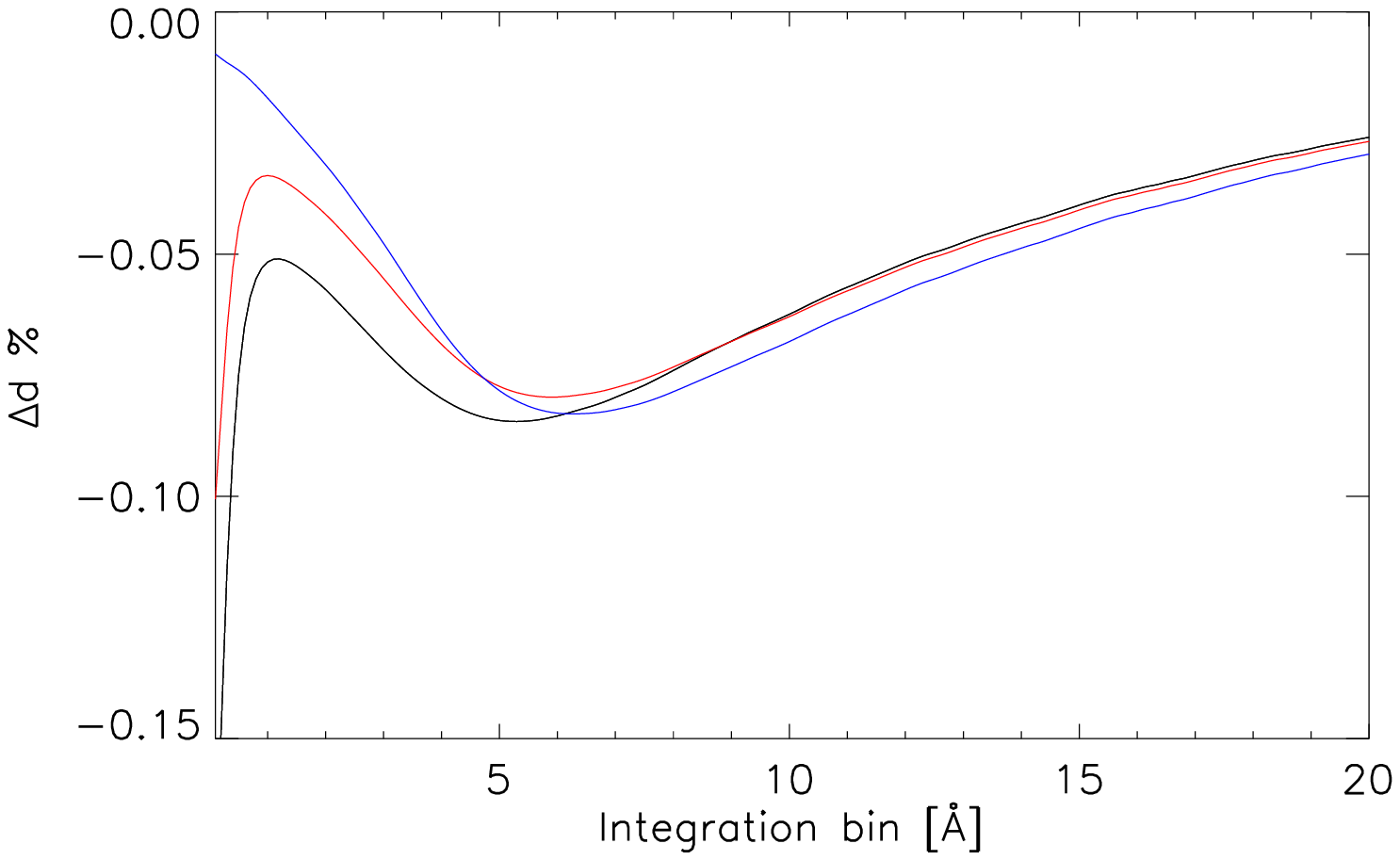}
\caption{$\Delta d$ as a function of integration bin for three different values of the ratio of the Mg{\sc ii}\,k to Mg{\sc ii}\,h line strength of the planetary atmospheric absorption, considering a Sun-like star and the default parameters. Black is for a line strength ratio of 2, that is close to what given by the \loggf\ values, red is for a line strength ratio of 1.5, while blue is for a line strength ratio of 1, that is equal line strength of the Mg{\sc ii}\,h\&k lines. The latter is close to what obtained from synthetic transmission spectra of KELT-9b computed accounting for NLTE effects \citep{fossati2021}.} \label{fig:figeratio}
\end{center}
\end{figure}

We study the impact of the considered Mg{\sc ii} planetary absorption on $\Delta d$. We computed $\Delta d$ considering both 5\% and 10\% peak absorption at the position of the Mg{\sc ii}\,k line and, for each of them, three planetary absorption line widths of FWHM equal to 0.75, 1.5, and 3.0\,\AA. These calculations have been done using the default parameters. Figure~\ref{fig:fig3} (top-left) shows that |$\Delta d$| increases with increasing planetary Mg {\sc ii} absorption. We also find that |$\Delta d$| initially decreases for very small integration bins and then increases with decreasing width of the planetary absorption features. Furthermore, as expected, the position of the second peak of |$\Delta d$| moves towards larger integration bins with increasing width of the planetary absorption features, demonstrating the tight relation between these two quantities. As mentioned in Section~\ref{sec:2-plparam}, our calculations assume that the planetary atmospheric gas responsible for the absorption is optically thin, which is often not the case. Therefore, we show in Figure~\ref{fig:figeratio} how $\Delta d$ varies as a function of integration bin for three different values of the ratio of the Mg{\sc ii}\,k to Mg{\sc ii}\,h line strength of the planetary atmospheric absorption, namely values of 2 (i.e. about what given by the \loggf\ values), 1.5, and 1 (i.e. equal line strength). Figure~\ref{fig:figeratio} indicates that the impact of line strength ratio is negligible compared to the size of the typical uncertainties of NUV transit depth measurements, and thus that the assumption of optically thin planetary atmospheric gas does not affect the general results. However, Figure~\ref{fig:figeratio} also shows that line strength ratio should be considered as a further free parameter for precisely computing $\Delta d$.

We further study the impact of stellar parameters on $\Delta d$. We compute $\Delta d$ considering three stars of spectral type K5 ($T_{\rm eff}$\,=\,4400\,K), G2 ($T_{\rm eff}$\,=\,5800\,K), and F5 ($T_{\rm eff}$\,=\,6500\,K) and, for each of them, we consider both 5\% and 10\% peak planetary absorption at the position of the Mg{\sc ii}\,k line. The results are shown in Figure~\ref{fig:fig3} (top-right). We find that stellar spectral type has a significant impact on the shape of $\Delta d$ as a function of the integration bin and in particular $\Delta d$ is on average larger for later spectral types. This is because the cooler the star the more important the chromospheric emission with respect to the photospheric emission. Therefore, the relative importance of the ISM absorption, which lies on top of the chromospheric emission that is absorbed by the planet, increases with decreasing stellar effective temperature. Furthermore, the top-right panel of Figure~\ref{fig:fig3} indicates that for cooler stars |$\Delta d$| decreases initially for very small integration bins and then increases with increasing integration bin, which would be against the expectation. However, for cooler stars |$\Delta d$| behaves as for hotter stars, but the position of the maximum is located at an integration bin that is beyond the shown range and then decreases at larger integration bins, as expected. Again, this stems out of the strong chromospheric emission that overwhelms the photospheric emission for K- and later type stars.

We also compute $\Delta d$ considering a Sun-like star, three different $E$ values of 50, 100, and 200\,erg\,cm$^{-2}$\,s$^{-1}$, and for each of them both 5\% and 10\% peak planetary absorption at the position of the Mg{\sc ii}\,k line. The results are shown in the middle-left panel of Figure~\ref{fig:fig3}, which indicates that |$\Delta d$| sharply decreases initially and then increases with increasing $E$ and also that the peak of |$\Delta d$| moves towards larger integration bins with increasing $E$.
\subsection{Impact of ISM parameters}\label{sec:ism_parameters}
The ISM absorption features have three main components that can affect $\Delta d$, namely $\log{N_{\rm MgII}}$, the $b$-parameter, and $RV_{\rm ISM}$. Employing the spectrographs on-board HST, \citet{redfield2002} measured these three parameters for a large number of stars lying within 100\,pc. Therefore, we consider the results of \citet{redfield2002} as a reference for varying the parameters describing the ISM absorption.

We computed $\Delta d$ setting the Mg{\sc ii} column density at 10$^{11}$, 3.2$\times$10$^{12}$, and 10$^{14}$\,cm$^{-2}$. For each of them, we further considered both 5\% and 10\% peak planetary absorption at the position of the Mg{\sc ii}\,k line. Figure~\ref{fig:fig3} (middle-right) shows that |$\Delta d$| increases with increasing $\log{N_{\rm MgII}}$. Similarly, we varied also the $b$-parameter between 1 and 5\,km\,s$^{-1}$ obtaining that |$\Delta d$| increases with increasing $b$ (bottom-left panel of Figure~\ref{fig:fig3}).

We also looked at the effect of varying $RV_{\rm ISM}$ keeping all other parameters constant at their default values, for the two considered peak values of the planetary absorption at the position of the Mg{\sc ii}\,k line (i.e. 5\% and 10\%). In particular, we varied $RV_{\rm ISM}$ between $-$50 and $+$50\,km\,s$^{-1}$ in steps of 25\,km\,s$^{-1}$. Because we model the synthetic chromospheric emission and ISM absorption features as symmetric features, the results obtained considering positive $RV_{\rm ISM}$ values are almost identical to those obtained considering negative $RV_{\rm ISM}$ values, where the differences are due to the fact that the photospheric stellar spectrum presents metal absorption lines around the core of the Mg{\sc ii} lines that are not symmetrically placed with respect to the Mg{\sc ii} line cores (e.g. top panel of Figure~\ref{fig:fig2}). However, these differences are negligible and we ignore them in the presentation of the results shown in the bottom-right panel of Figure~\ref{fig:fig3}. We find that |$\Delta d$| is larger for smaller $RV_{\rm ISM}$ values, namely where the ISM absorption features lie closer to the maximum of the stellar emission. For large $RV_{\rm ISM}$ values and small integration bins, $\Delta d$ varies significantly as a function of bin size. This is due to the fact that for $RV_{\rm ISM}$ values of $\pm$30--50\,km\,s$^{-1}$ other photospheric lines begin to play a role, while the ISM absorption features are located close to the minimum of the stellar emission. 
\subsection{Impact of emission width}\label{sec:emission_width}
As mentioned in Section~\ref{sec:params}, the width of the chromospheric Mg{\sc ii} stellar emission varies as a function of stellar type and luminosity. Thus we further check whether this plays a role in the shape of $\Delta d$ as a function of integration bin. To this end, we further derived the width of the Mg{\sc ii}\,h\&k line emission from an HST/STIS observation of the early K-type star $\epsilon$\,Eri obtaining values of 0.394 and 0.431\,\AA, respectively, which are slightly lower than those we measured for $\alpha$\,Cen\,A, which is an early G-type star. 

\begin{figure}
\begin{center}
\includegraphics[width=\hsize]{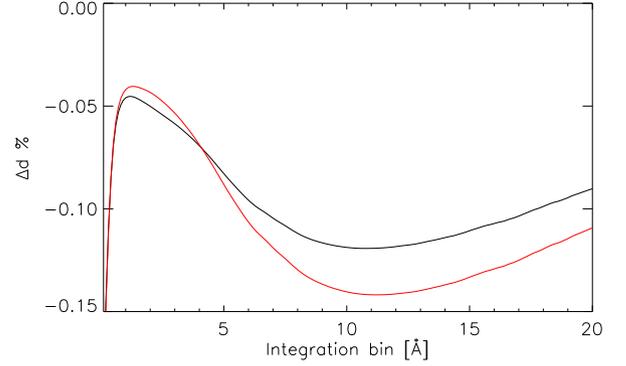}
\caption{$\Delta d$ as a function of integration bin for two different sets of Mg{\sc ii}\,h\&k line emission widths, all other parameters being as default. The black line is for the set of line widths obtained from the spectrum of $\alpha$\,Cen\,A, while the red line is for the set of smaller line widths obtained from the spectrum of $\epsilon$\,Eri.}
\label{fig:figewidth}
\end{center}
\end{figure}

Figure~\ref{fig:figewidth} shows how $\Delta d$ varies as a function of integration bin considering the stellar parameters (Temperature: 4975\,K; Radius: 0.83\,$R_{\odot}$) of $\epsilon$\,Eri given by \cite{gaiadr2} and the two set of chromospheric line emission widths (i.e. as obtained from the spectra of $\alpha$\,Cen\,A and $\epsilon$\,Eri), all other parameters being as the default ones. On average, we find larger |$\Delta d$| values considering smaller chromospheric line widths, but in general the impact of the stellar emission line width on |$\Delta d$| is small, particularly when compared to the typical uncertainties of NUV transit depth measurements.
%
\section{Discussion}\label{sec:discussion}
Figures~\ref{fig:fig5} to \ref{fig:fig10} show $\Delta d$ as a function of $\log{N_{\rm MgII}}$ (ranging between 10$^{10}$ and 3.2$\times$10$^{15}$\,cm$^{-2}$) and $RV_{\rm ISM}$ (ranging between $-$50 and $+$50\,km\,s$^{-1}$), and for the two peak values of the planetary absorption at the position of the Mg{\sc ii}\,k line of 5\% and 10\% we consider in this work. In each plot, we further individually vary spectral resolution (Figure~\ref{fig:fig5}), integration bin (Figure~\ref{fig:fig6}), FWHM of the planetary absorption (Figure~\ref{fig:fig7}), stellar spectral type (Figure~\ref{fig:fig8}), $E$ value (Figure~\ref{fig:fig9}), or $b$-parameter (Figure~\ref{fig:fig10}). For all cases and unless specified, we consider a Sun-like star and fix $E$ equal to 100\,erg\,cm$^{-2}$\,s$^{-1}$, the $b$-parameter equal to 3\,km\,s$^{-1}$, $R$ equal to 30000, the FWHM of the planetary absorption features equal to 1.5\,\AA, and an integration bin equal to 5.0\,\AA.

The maximum of |$\Delta d$| shown in these plots is about 0.004, which is comparable to the uncertainties of previous NUV transmission spectroscopy measurements presented by \citet[][WASP-12b]{fossati2010}, \citet[][WASP-121b]{sing2019}, and \citet[][HD209458b]{cubillos2020}, in the region covered by the Mg{\sc ii}\,h\&k lines (though the integration bins are different in the three works). However, the average value of |$\Delta d$| is of the order of 0.001 which is smaller than the typical uncertainties of the NUV transmission spectroscopy observations published so far. We remark that the properties of the ISM absorption we considered are realistic, since they are based on results of ISM absorption measurements for stars within 100\,pc. 

In the following, we present an open-source tool enabling one to estimate $\Delta d$ on the basis of parameters given by the user and the algorithm described in Section~\ref{sec:params} and \ref{sec:analysis}. We also present the $\Delta d$ value obtained employing this tool for a few systems observed, or soon to be observed, in the NUV.
\subsection{Correcting for unresolved Mg{\sc ii} ISM absorption in transit depth measurements}\label{sec:correction}
As shown in the previous sections, estimating $\Delta d$ requires one knowing the shape of the planetary transmission spectrum, of the ISM absorption features, and of the stellar lines. Therefore, it is not possible to accurately evaluate $\Delta d$, particularly in the cases when the ISM absorption features and/or the stellar Mg{\sc ii}\,h\&k lines are not well resolved. Estimating $\Delta d$ requires models of the planetary transmission spectrum and possibly estimates of the ISM absorption and Mg{\sc ii} emission features based mostly on scaling relations. Furthermore, once these ingredients are known, one has to evaluate their impact on $\Delta d$ considering the complicated interplay among the many parameters involved. To aid in this task, we developed an open-source tool\footnote{https://sites.google.com/view/geco-exoplanets/home/software?authuser=0}, available both in Python and IDL, enabling one to estimate $\Delta d$ following the recipes presented above and scaling relations for estimating the Mg{\sc ii}\,h\&k line core emission and ISM absorption that we describe below. 

The code requires as input the stellar effective temperature $T_{\rm eff}$ (in K), stellar radius (in $R_{\odot}$), disk-integrated Mg{\sc ii}\,h\&k line core emission at the distance of 1\,AU (i.e. $E$ in erg\,cm$^{-2}$\,s$^{-1}$), the broad-band (i.e. continuum) NUV transit depth, the peak transit depth at the position of the Mg{\sc ii}\,k line, the width of the planetary absorption features (in \AA), $\log{N_{\rm MgII}}$, $RV_{\rm ISM}$ (in km\,s$^{-1}$), the instrumental spectral resolution $R$ (in \AA), and the size of the integration bin (in \AA). Based on the results of \citet{redfield2002}, we fix the $b$-parameter equal to 3\,km\,s$^{-1}$. To run the code, the user has to provide as input at least $T_{\rm eff}$, the broad-band (i.e. continuum) NUV transit depth, the peak of the transit depth at the position of the Mg{\sc ii}\,k line, the width of the planetary absorption features, $R$, and the size of the integration bin. If unavailable, the other parameters are derived by the code from assumptions and scaling relations. If not given by the user, the stellar radius is obtained by interpolating $T_{\rm eff}$ on Table~5 of \citet{pecaut2013}. Instead, the code derives the Mg{\sc ii} chromospheric emission on the basis of the \logR\ value (given by the user as an extra input parameter) and the ISM parameters on the basis of literature results and interstellar reddening (given by the user as an extra input parameter), as described below. 
\subsubsection{Mg\,{\sc ii}\,h\&k line core emission from \logR\ measurements}\label{sec:mg2emission}
The \logR\ parameter is a measure of the chromospheric emission in the core of the Ca{\sc ii}\,H\&K lines and as such, once appropriately calibrated, can be used to estimate the chromospheric emission in the core of the Mg{\sc ii}\,h\&k lines. We employed \logR\ measurements collected by \citet[][and references therein]{sreejith2020} and Mg{\sc ii}\,h\&k line core emission measurements given by \citet{linsky2013} to develop scaling relations between the two quantities as a function of stellar spectral type. To this end, we follow the same approach described by \citet{sreejith2020} and find that $\log{E}$ and \logR\ are linearly correlated as
\begin{equation}
\label{eq:our_result}
    \log_{10}{E}=c_1 \times \logR\ +c_2\,,
\end{equation}
where the $c_1$ and $c_2$ coefficients are listed in Table~\ref{table:2}.
\begin{figure}
\begin{center}
\includegraphics[width=\hsize]{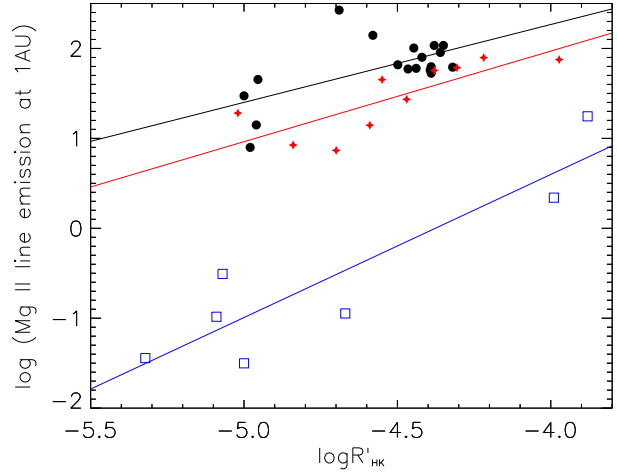}
\caption{Correlation between Ca{\sc ii}\,H\&K stellar activity index (\logR) and $\log{E}$ for F-, G- (black dots), K- (red star), M-type stars (open blue squares). The corresponding linear fits are represented with the same colour as the symbols.}
\label{fig:fig4}
\end{center}
\end{figure}
\begin{table}
\caption{Correlation parameters of the linear fits between stellar activity index (\logR) and $\log{E}$ defined in Equation~(\ref{eq:our_result}).}     
\label{table:2}
\centering                                      
\begin{tabular}{c c c }          
\hline\hline                        
Sp. Type & c$_1$ & c$_2$ \\    
\hline                                  
    F, G &  0.87 & 5.73   \\ 
    K &  1.01 & 6.00       \\
    M & 1.59 & 6.96        \\
 \hline                                            
\end{tabular}
\end{table}

If the user does not insert the $E$ value in the code, then the code requests the \logR\ parameter to compute $E$ as described above. Also, the code selects the spectral type required to select the parameters to use for the calibration (Table~\ref{table:1}) on the basis of $T_{\rm eff}$ and Table 5 of \citet{pecaut2013}.

\subsubsection{Deriving Mg\,{\sc ii} ISM parameters}\label{sec:ism}
If the user does not insert the Mg{\sc ii} ISM parameters, the code assumes $RV_{\rm ISM}$ to be equal to 0\,km\,s$^{-1}$ and the $b$-parameter to be equal to 3\,km\,s$^{-1}$, while $\log{N_{\rm MgII}}$ is derived from literature results as follows. The Mg{\sc ii} ISM column density is computed employing the Mg ISM abundance and ionisation fraction given by \citet{frisch} (their Table~5) as
\begin{equation}
N_{\rm MgII} = N_{\rm H} \times {\rm Mg~ISM~abundance} \times {\rm MgII~ionisation~fraction}\,,
\end{equation}
where \citep{savage}
\begin{equation}
N_{\rm H} = 5.8 \times 10^{21}\,E(B-V)
\end{equation}
is the total hydrogen column density (i.e. H$_2+$H{\sc i}$+$H{\sc ii} atoms\,cm$^{-2}$) as a function of the stellar reddening $E(B-V)$, which has to be provided by the user. 

\subsection{Relevance for exoplanet studies}
We apply the code described above to a few specific systems that have been observed in the NUV or are soon going to be. The aim is to compare $\Delta d$ with the observational uncertainties to identify the potential impact of ISM absorption on transit depth measurements and thus on transmission spectra. In particular, we look at WASP-12b \citep{fossati2010,haswell2012}, WASP-121b \citep{sing2019}, HD209458b \citep{cubillos2020}, which have been observed with HST, and KELT-9b, which is going to be one of the first targets that is going to be observed by the CUTE SmallSat mission \citep{fleming2018}.

WASP-12b \citep{hebb2009} has been the first exoplanet to be observed at NUV wavelengths \citep{fossati2010,haswell2012}. We estimated the impact of Mg{\sc ii} ISM absorption on the transit depths measured by \citet{fossati2010} and \citet{haswell2012} considering the stellar parameters given by GAIA \citep{gaiadr2}. Following the discussion of \citet{fossati2013}, we estimated the Mg{\sc ii} chromospheric emission employing a \logR\ value of $-$5.1 and the conversion presented in Section~\ref{sec:mg2emission}. We used the bluest value in the optical transmission spectrum presented by \citet{Sing2016} as the broadband transit depth. The peak transit depth at the position of the Mg\,{\sc ii}\,k line was estimated on the basis of the transit depth measured by \citet{fossati2010} and considering that the measurement has been obtained using an integration bin of about 20\,\AA. We further assumed a width of the planetary absorption feature of 1.5\,\AA, following what obtained by \citet{sing2019} for WASP-121b. The ISM parameters were derived as described in Section~\ref{sec:ism} and employing the reddening inferred from the ISM absorption maps of \citet{amores2005} and the GAIA distance. For an integration bin of 20\,\AA\ corresponding to that of the observations, we find a $\Delta d$ value of $-1.46\times10^{-4}$, which is well within the measurement uncertainties \citep[$\approx$0.5\%;][]{fossati2010,haswell2012}.

WASP-121b is a transiting hot Jupiter, similar to WASP-12b, with an orbital period of 1.27\,days \citep{evans2017}. \citet{sing2019} analysed NUV HST/STIS transit observations covering the Mg{\sc ii}\,h\&k lines detecting the presence of both Mg and Fe in the upper planetary atmosphere. We estimated the stellar, chromospheric emission, and ISM absorption parameters as we have done for WASP-12b, while we considered the strength of the planetary absorption observed by \citet{sing2019}. At the resolution of the STIS spectrograph and for the integration bin of 10\,\AA\ employed by \citet[][their Figure~12]{sing2019}, we find a $\Delta d$ value of $-1.51\times10^{-3}$, which suggests that the measured Mg{\sc ii} transit depth may have been slightly underestimated, though the obtained $\Delta d$ value is within the 1$\sigma$ observational uncertainties.

HD209458b was the first discovered transiting planet  \citep{Charbonneau2000,henry2000} and one of the most well studied planets with several species detected in its atmosphere \citep[e.g.][]{giacobbe2021}. \citet{cubillos2020} reanalysed the NUV HST/STIS transmission spectroscopy observations of HD209458b originally presented by \citet{vidal2013} detecting Fe, but not Mg, in the planetary upper atmosphere, suggesting that the planet hosts Mg-bearing aerosols in the lower atmosphere. We obtained the stellar parameters from GAIA \citep{gaiadr2} and employed a \logR\ value of $-$4.97 to infer the Mg{\sc ii} chromospheric emission. We estimated the Mg{\sc ii} ISM column density from the H{\sc i} ISM column density measured by \citet{wood2005} and the scaling relations given in Section\ref{sec:ism}. As for WASP-12b, we used the bluest value in the optical transmission spectrum presented by \citet{Sing2016} as the broadband transit depth. Because of the Mg{\sc ii} non-detection, we considered peaks of the planetary absorption at the position of the Mg{\sc ii}\,k line smaller than 5\% and a width of 1.5\,\AA. We find a $\Delta d$ value of 0.023 for an integration bin of 0.23\,\AA\ and of 4.33$\times10^{-3}$ for an integration bin of 0.93\,\AA\ \citep[see the top and bottom panels of Figure~5 of][]{cubillos2020}. For HD209568b, the weaker ISM absorption due to the closer distance to the star leads to positive $\Delta d$ values indicating that ISM absorption would in principle lead to overestimate the transit depth. However, \citet{cubillos2020} did not find any planetary absorption signal and thus our results strengthen the non-detection as ISM absorption would not be able to hide the planetary absorption signal in this case.

KELT-9b is an ultra-hot Jupiter orbiting an early A-type star every 1.5 days \citep{gaudi2017}. As a result of its high atmospheric temperature and bright host star, this is one of the currently most studied planets and a high-priority target for the CUTE SmallSat. We obtained the stellar parameters and estimated the ISM parameters as for WASP-12b. We assumed a broadband transit depth equal to that obtained in the optical \citep{gaudi2017} and considered a peak value of the planetary absorption at the position of the Mg{\sc ii}\,k line of 10\%. For an integration bin of 5\,\AA\ and at the resolution of HST/STIS and CUTE, we find $\Delta d$ values of $-1. 2\times10^{-5}$ and $-2.2\times 10^{-5}$, respectively, that is significantly smaller than the uncertainties one expects to obtain from the observations \citep[e.g.][]{sreejith2019}.

\section{Conclusion}\label{sec:conclusion}

We studied and quantified the impact of unresolved or unconsidered Mg{\sc ii} ISM absorption in the core of the Mg{\sc ii}\,h\&k lines on transit depth measurements. We carried out this work employing synthetic spectra, varying a number of stellar, planetary, ISM, instrumental, and analysis parameters. The bias induced by ignoring ISM absorption might lead to overestimate or underestimate transit depths, depending on the specific input parameters. Our parameter study showed that in general |$\Delta d$| (i.e. the difference in transit depth obtained considering ISM absorption and ignoring it) increases with decreasing stellar temperature, increasing Mg{\sc ii} chromospheric emission, increasing planetary absorption, decreasing planetary absorption width, increasing ISM absorption, and increasing $b$-parameter. |$\Delta d$| is also maximum when the ISM absorption and chromospheric emission features overlap (i.e. $RV_{\rm ISM}$ equal to 0\,km\,s$^{-1}$). We also find that the integration bin considered for computing the transit depth plays a key role in determining the value of $\Delta d$, with $\Delta d$ typically decreasing with increasing integration width.

From all runs we carried out, we found maximum $\Delta d$ values of the order of 0.005--0.01, though the actual values depend strongly on the involved parameters. The average $\Delta d$ values we obtain are of the order of 0.0005--0.001, which is smaller than the uncertainties typically obtained with the currently available facilities. As an example, we estimated $\Delta d$ for WASP-12b, WASP-121b, HD209458b, and KELT-9b, which have been or are soon going to be observed in the NUV. In all cases, we obtained values smaller than the observational uncertainties. However, this may not be the case of observations carried out employing future, larger facilities (e.g. HABEX, LUVOIR) for which the observational uncertainties would be significantly smaller. 

\section*{Acknowledgements}

A.G.S. and L.F. acknowledge financial support from the Austrian Forschungsf\"orderungsgesellschaft FFG projects CONTROL P865968 and CARNIVALS P885348. This project was also partly funded by the Austrian Science Fund (FWF) [J 4596-N]. P.C. acknowledges support and funding by the Austrian Science Fund (FWF) Erwin Schroedinger Fellowship program J4595-N. We thank the anonymous referee for their comments that helped to significantly improve the paper.

\section*{Data Availability}
The data underlying this article will be shared on reasonable request to the corresponding author.
 



\bibliographystyle{mnras}
\bibliography{example} 


\appendix

\section{}

\begin{figure*}
\begin{center}
\resizebox{\hsize}{!}
{\includegraphics[width=\textwidth]{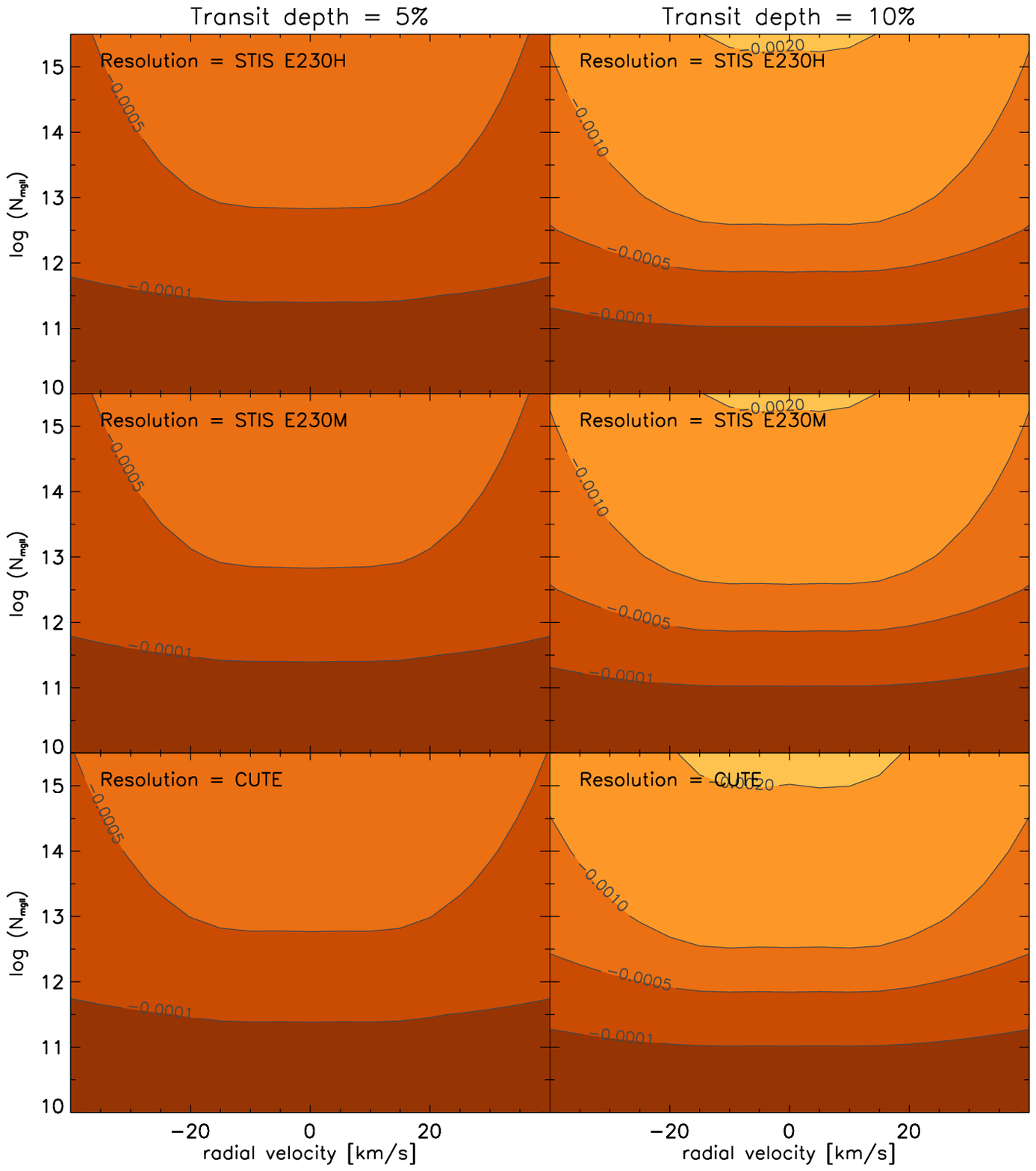}}
\caption{$\Delta d$ (contours) as a function of $RV_{\rm ISM}$ and $\log{N_{\rm MgII}}$ for two different peak values of the planetary atmospheric absorption at the position of the Mg{\sc ii}\,k line of 5\% (left) and 10\% (right) and at the three considered spectral resolutions. We assumed a Sun-like star (i.e. $T_{\rm eff}$\,=\,5800\,K; $R_{\rm s}$\,=\,1\,$R_{\odot}$), $E$\,=\,100\,erg\,cm$^{-2}$\,s$^{-1}$, $b$\,=\,3\,km\,s$^{-1}$, a width of the planetary absorption features of 1.5\,\AA, and an integration bin of 5\,\AA.}  
\label{fig:fig5}
\end{center}
\end{figure*}

\begin{figure*}
\begin{center}
\resizebox{\hsize}{!}
{\includegraphics[width=\textwidth]{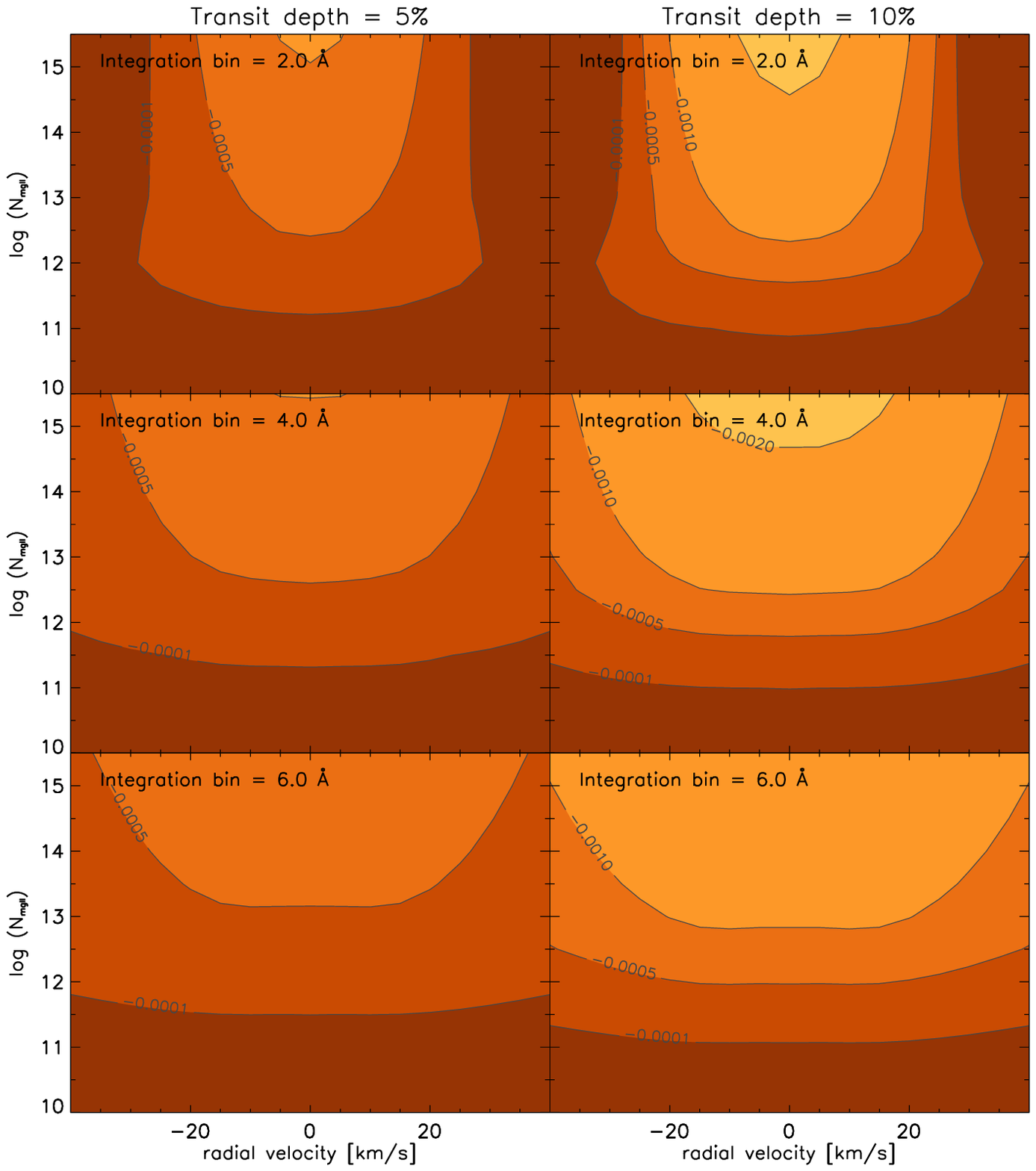}}
\caption{Same as Figure~\ref{fig:fig5}, but varying integration bin and at a fixed spectral resolution of 30000.}
\label{fig:fig6}
\end{center}
\end{figure*}

\begin{figure*}
\begin{center}
\resizebox{\hsize}{!}
{\includegraphics[width=\textwidth]{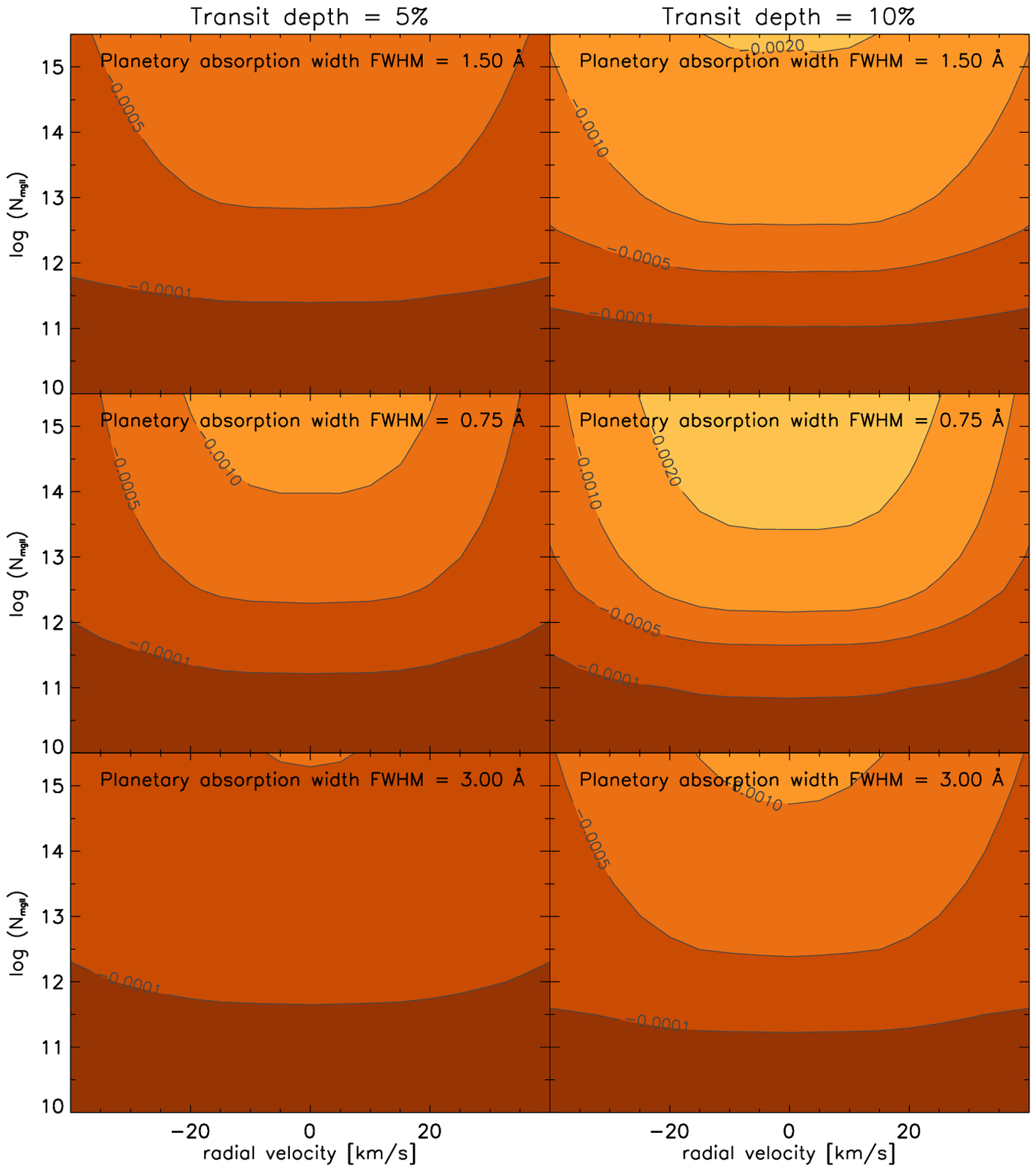}}
\caption{Same as Figure~\ref{fig:fig5}, but varying the width of the planetary absorption features and at a fixed spectral resolution of 30000.}
\label{fig:fig7}
\end{center}
\end{figure*}

\begin{figure*}
\begin{center}
\resizebox{\hsize}{!}
{\includegraphics[width=\textwidth]{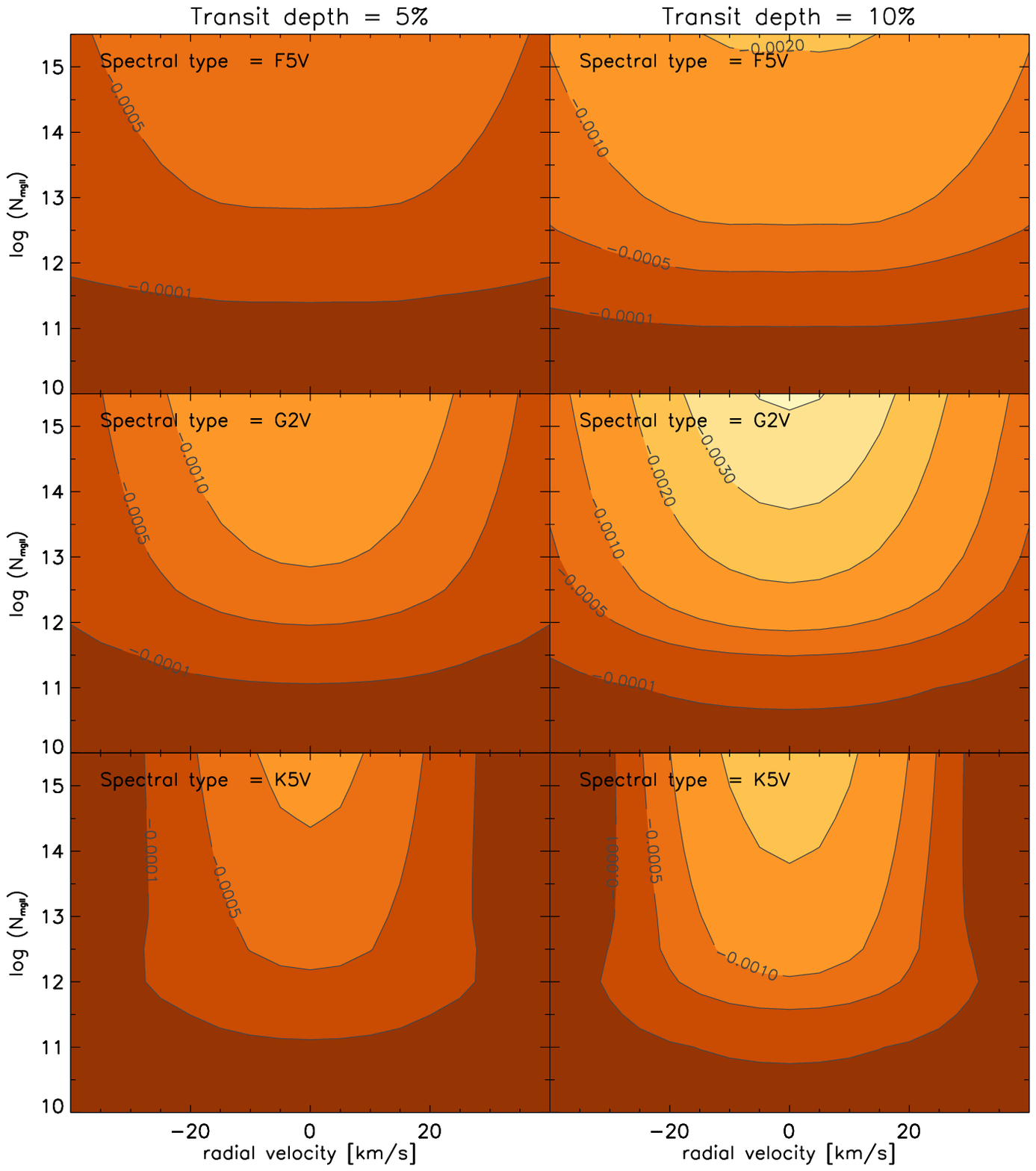}}
\caption{Same as Figure~\ref{fig:fig5}, but varying the stellar spectral type and at a fixed spectral resolution of 30000.}
\label{fig:fig8}
\end{center}
\end{figure*}

\begin{figure*}
\begin{center}
\resizebox{\hsize}{!}
{\includegraphics[width=\textwidth]{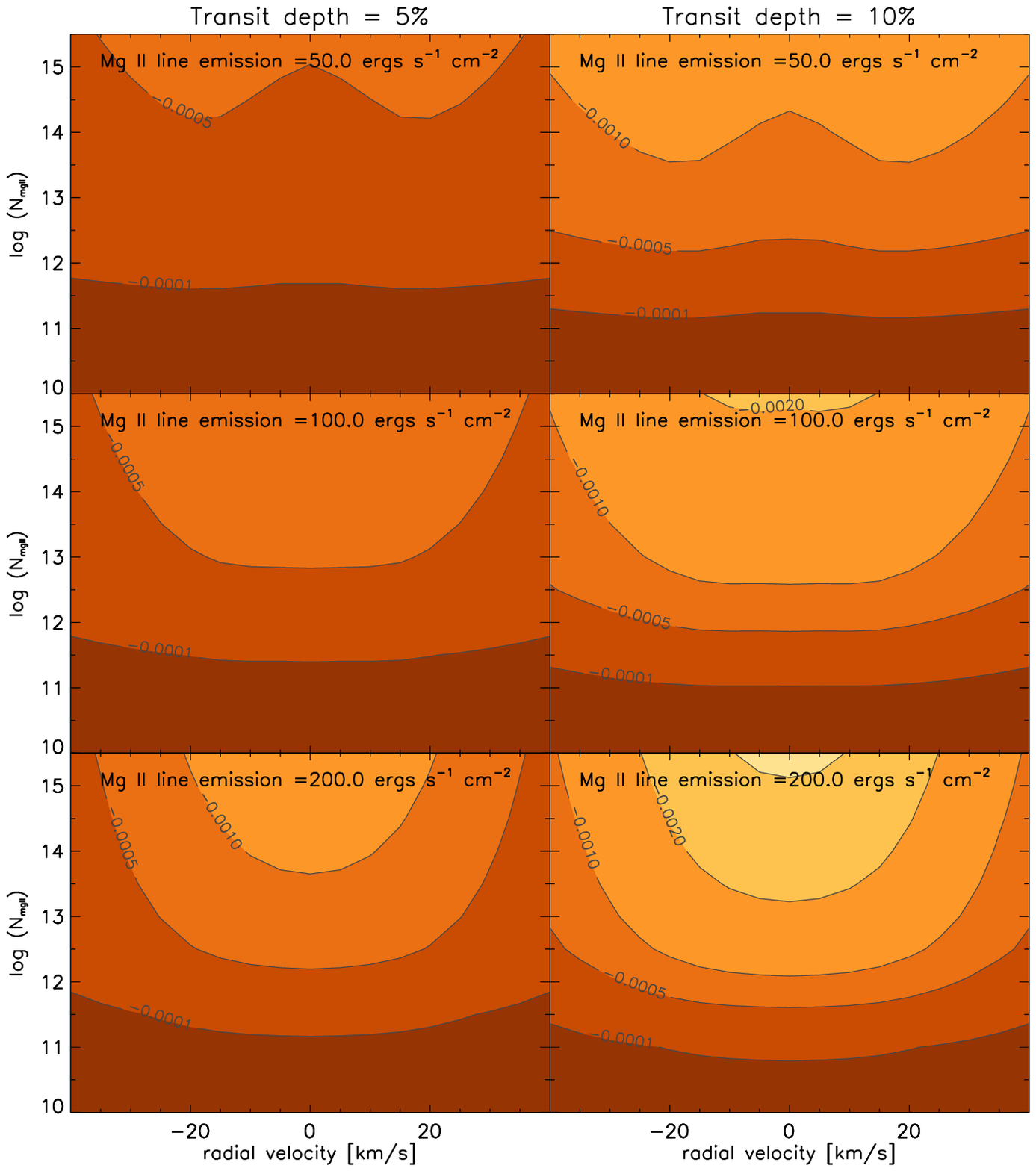}}
\caption{Same as Figure~\ref{fig:fig5}, but varying the stellar Mg{\sc ii} chromospheric emission and at a fixed spectral resolution of 30000.}
\label{fig:fig9}
\end{center}
\end{figure*}

\begin{figure*}
\begin{center}
{\includegraphics[width=\textwidth]{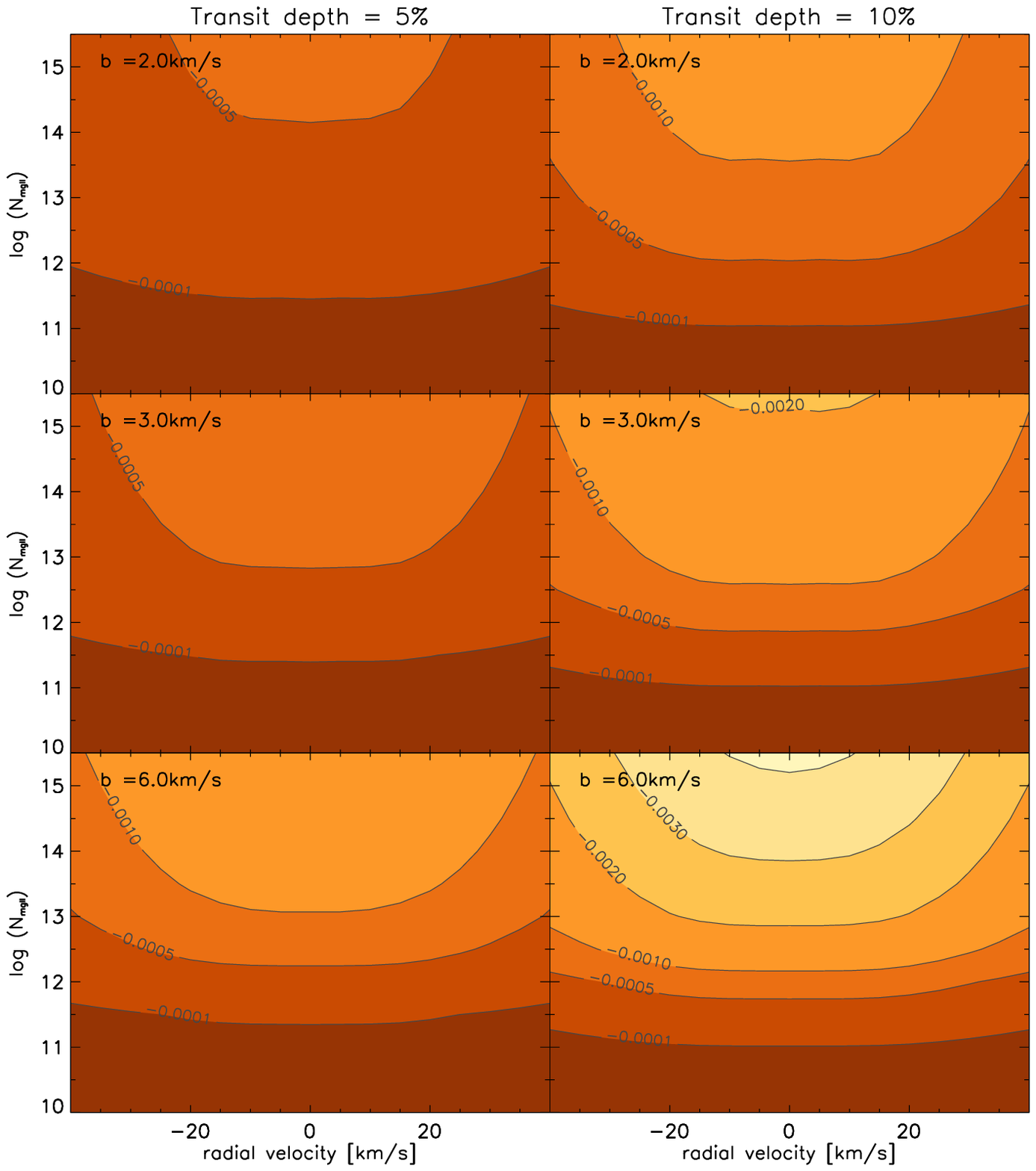}}
\caption{Same as Figure~\ref{fig:fig5}, but varying the ISM broadening $b$-parameter and at a fixed spectral resolution of 30000.}
\label{fig:fig10}
\end{center}
\end{figure*}


\bsp	
\label{lastpage}
\end{document}